\documentclass[aps,prl,reprint,amsmath,amssymb,superscriptaddress,nofootinbib,longbibliography,floatfix,flushbottom,balancelastpage]{revtex4-2}
\usepackage[T1]{fontenc}
\usepackage{lmodern}
\usepackage{graphicx,bm,booktabs}
\usepackage[colorlinks=true,linkcolor=blue,citecolor=blue,urlcolor=blue]{hyperref}
\newcommand{\dd}{\mathrm{d}}
\newcommand{\tr}{\operatorname{tr}}
\newcommand{\diag}{\operatorname{diag}}
\newcommand{\sech}{\operatorname{sech}}
\newcommand{\PT}{P_{\mathrm T}}
\newcommand{\calK}{\mathcal K}
\newcommand{\calH}{\mathcal H}

\newcommand{\calS}{\mathcal S}
\newcommand{\calE}{\mathcal E}
\newcommand{\calB}{\mathcal B}
\newcommand{\pzero}{p_{\psi,0}}
\begin{document}
\title{Coherent Selection of Critical Kasner Response}
\author{Yi-kun Li}
\affiliation{State Key Laboratory of Radio Astronomy and Technology,
Xinjiang Astronomical Observatory, CAS, 150 Science 1-Street,
Urumqi 830011, China}
\affiliation{College of Astronomy and Space Science,
University of Chinese Academy of Sciences, No.1 Yanqihu East Road,
Beijing 101408, China}
\date[Correspondence: ]{\texttt{liyikun@xao.ac.cn}}
\begin{abstract}
A coherent wave component with extremely small initial self-energy can substantially change the first response of a contracting spacetime. We demonstrate this effect in Einstein--Maxwell--scalar theory with two commuting spacelike Killing fields. Along a critical family motivated by scalarized black-hole interiors, the estimated magnetic delay leaves time for longitudinal propagation to redistribute the incident field. Increasing a low-frequency component at fixed initial electromagnetic energy enhances the scalar response on one background and suppresses it on another. A calculation that propagates the field and its momentum, then their quadratic scalar and geometric sources, predicts this reversal before nonlinear evolution. At the first suppression comparison, the coherent profile error is $0.38\%$, against $46.9\%$ when cross terms are omitted. The selected response develops into a magnetic pulse with electric and scalar exchange. The critical geometry thus preserves incident coherence as information that predicts its first departure from a Kasner epoch.
\end{abstract}
\maketitle
\emph{Introduction.}
Near a spacelike singularity, the geometry can contract at different rates in different directions. During a Kasner epoch these rates, normalized by the overall contraction, are nearly constant. Matter can grow during such an epoch and redirect the evolution through a brief interaction, represented by a potential wall~\cite{BKL1970,DamourHenneauxNicolai2003}. Numerical studies reveal both sequences of local transitions and spatial structures that influence them~\cite{Garfinkle2004,Andersson2005}. An incident wave brings a further physical ingredient: its electric and magnetic content changes as it propagates. Understanding the first response then requires knowing which information about the incoming wave survives to affect the contracting geometry.

The interiors of scalarized charged black holes supply a critical family in which this propagation can be studied~\cite{LiSunYang2026}. Near the scalarization threshold, two spatial Kasner exponents and the scalar velocity approach zero. The estimated magnetic response is delayed, while propagation along the remaining direction stays effective. A weak field therefore has time to change its profile before appreciably affecting the scalar motion. Electromagnetic amplification by a scalar coupling is familiar~\cite{Sobol2018}. Here the critical geometry determines the competing propagation and response times. Together with a prescribed physical packet duration and the carrier's complex transmission, it specifies how the response can change between backgrounds.

We study this mechanism in a local Einstein--Maxwell--scalar (EMS) model with two commuting spacelike Killing fields. The field propagates in one spatial direction and exchanges energy with the scalar and geometry. We find that increasing a coherent low-frequency component at fixed initial electromagnetic energy can enhance or suppress the first scalar response. Its effect reverses along the specified background family. Propagating the field and its momentum predicts both the reversed ordering and the early spatial response, measured by the scalar velocity relative to the contraction rate. The ensuing magnetic pulse develops into nonlinear electric and scalar exchange. Thus incoming coherence selects a measurable departure from an almost free Kasner epoch.

\emph{Critical geometry and propagation.}
We use $G=c=1$, signature $(-,+,+,+)$, and the action
\begin{equation}
 I=\frac{1}{16\pi}\int\dd^4x\sqrt{-g}
 \left[R-2(\nabla\psi)^2-e^{\psi^2}F_{\mu\nu}F^{\mu\nu}\right].
 \label{eq:action}
\end{equation}
The local contraction rates are encoded in the extrinsic curvature. On the contracting foliation, let $n$ point toward contraction and $K_{ij}=\mathcal L_n\gamma_{ij}/2$. Dividing by the total contraction gives the geometric spectrum
\begin{equation}
 P=\left(\operatorname{eig}\frac{K^i{}_j}{\tr K};p_\psi\right),
 \qquad p_\psi=\frac{\sqrt2\,n^\mu\nabla_\mu\psi}{|\tr K|}.
 \label{eq:spectrum}
\end{equation}
Its spatial entries measure the relative directional rates, and its scalar entry measures the scalar velocity in the same contraction units. The scalar $\varphi=\sqrt2\psi$ obeys the free Kasner constraint $\sum p_a^2+p_\psi^2=1$, with $\sum p_a=1$. Only spatial eigenvalues are permuted in comparisons.

Write $\delta=(Q_{\rm BH}/M)/q_c-1$ for the stationary family, with $q_c\simeq0.994$~\cite{LiSunYang2026}. Its interior parameter satisfies $\beta\sqrt\delta\to c_0\simeq0.07682$, with $p_s=2/(\beta^2+3)$, $p_\parallel=(\beta^2-1)/(\beta^2+3)$ and $p_\psi=2\sqrt2\beta/(\beta^2+3)$. Hence $p_s=O(\delta)$, the scalar velocity $v=p_\psi/\sqrt2=O(\sqrt\delta)$, and $P\to(0,0,1;0)$. The simulations use the measured exponents at each finite background.
The two equal exponents $p_s$ describe the transverse directions; $p_\parallel$ singles out the direction retained in the propagation problem.

A free-flight estimate makes the relevant timescale explicit. At logarithmic volume time $t$ after an incoming surface, take $\psi=\psi_K+vt$ and a frozen magnetic amplitude $\mathcal A_B$. Define $L_B=\ln|\mathcal A_B|^{-1}$ and the wall proxy
\begin{equation}
 \ln\mathcal R_B=-2L_B-2p_st+(\psi_K+vt)^2-\psi_K^2.
 \label{eq:clock}
\end{equation}
For an algebraically vanishing seed and bounded $\psi_K$, the threshold $\mathcal R_B(t_*)=1$ gives $t_*\sim(c_0/\sqrt2)\sqrt{L_B/\delta}$. At this estimated time the normalized longitudinal frame behaves as
\begin{equation}
 e^{-(1-p_\parallel)t_*}
 =\exp[-O(\sqrt{\delta L_B})]\longrightarrow1.
 \label{eq:frame}
\end{equation}
The transverse frame factors are exponentially small in $\sqrt{L_B/\delta}$. Longitudinal transport therefore remains effective during this estimated delay. The actual magnetic maximum depends additionally on its vanishing peak energy, whose contribution to the clock is discussed in the Supplemental Material~\cite{Supplement}.

At fixed positive $\delta$, the longitudinal frame eventually decays and suppresses transport. Approaching the critical state makes that decay slower than the estimated magnetic response. The disturbance can consequently redistribute its electric and magnetic content while the geometry remains close to its incoming epoch. The limiting Kasner exponents alone do not describe this intervening field evolution. The frame estimate identifies the surviving propagation direction; its effect on the scalar and geometry follows from the wave equations.

The two symmetry directions form surfaces whose area and shape can evolve. Use conformal coordinates $(T,y)$ for time and the remaining spatial direction, with base metric $e^{2\sigma}(-\dd T^2+\dd y^2)$ and symmetry-surface area $\rho=e^r$. The orbit connection describes the metric coupling between these surfaces and the base. The experiments occupy the consistent sector $F_{AB}=0$, with one transverse Maxwell potential $a_1$ and one orbit-shape polarization. Variation before fixing the base metric retains both local constraints. Solving them includes the initial geometry induced by the packet's energy and momentum flux~\cite{Supplement}.

If $\Lambda=a_\parallel|\tr K|$ converts the physical longitudinal coordinate to $y$ initially, a carrier of frequency $\omega$ and duration $M\tau_0/\sqrt\delta$ has
\begin{equation}
 k=\omega/\Lambda,\qquad w=\Lambda M\tau_0/\sqrt\delta,
 \qquad kw=\frac{(\omega M)\tau_0}{\sqrt\delta}.
 \label{eq:kw}
\end{equation}
The half-width $w$ sets a residence time since the base characteristics have speeds $\pm1$. The product $kw$ measures the carrier's phase variation across that width. Changing $\delta$ therefore changes how many oscillations the packet contains, together with the background and transmitted phase.

\emph{Coherent response.}
Expand about the homogeneous solution of the same charged model with no packet. Denote its fields by $r_0,\sigma_0,\psi_0$ and its conserved electric momentum by $q$. For $U=e^{\psi_0^2/2}a_1$, $g=\psi_0\psi_{0,T}$ and $b=(q^2/8)e^{-\psi_0^2-2r_0+2\sigma_0}$, linear propagation obeys
\begin{align}
 U_{TT}-U_{yy}&=(g_T+g^2-2b)U,\notag\\
 e&=U_T-gU,\qquad m=U_y.
 \label{eq:linear}
\end{align}
Here $e,m$ are the rescaled electric and magnetic fields. The changing scalar coupling supplies the time-dependent pump $g_T+g^2$; the orbit connection and background electric momentum supply the restoring term $-2b$. Amplification depends on the field's entire passage through this evolving pump.

Each wave mode needs an initial amplitude and velocity, just as an oscillator needs its initial position and momentum. Their relative phase sets the subsequent electric and magnetic fields even at fixed mode energy. A compact packet combines a continuous band of such modes. The low-frequency component and carrier evolve differently in the pump, changing their overlap in the region where the scalar responds. Both the incoming phase and the phase accumulated during propagation enter this overlap.

Electric and magnetic fields push the scalar in opposite directions through the signed source $\calS=e^2-m^2$. Propagating this source and the accompanying $U^2$ terms gives the second-order changes $(R,\Sigma,D,\Phi)$ of the orbit area, base metric, orbit shape and scalar, respectively. The predicted geometric response is
\begin{equation}
 \Delta p_\psi^{(2)}
 =\frac{\sqrt2}{|\calK_0|}
 \left[\Phi_T-\frac{\psi_{0,T}}{\calK_0}(R_T+\Sigma_T)\right],
 \label{eq:response}
\end{equation}
where $\calK_0=r_{0,T}+\sigma_{0,T}<0$. The first term measures the change in scalar velocity; the second includes the change in the contraction rate used to normalize it. The forced equations and their constrained initial data are given in End Matter.

For a carrier and a smooth low-frequency addition, $U=N(U_c+\lambda U_\ell)$, the signed source contains $2\lambda(e_ce_\ell-m_cm_\ell)$ as well as both self-terms. Consequently
\begin{equation}
 \Delta p_\psi^{(2)}=N^2
 \left(C_{cc}+\lambda C_{c\ell}+\lambda^2C_{\ell\ell}\right).
 \label{eq:quadratic}
\end{equation}
The bilinear coefficients include propagation, accumulated scalar exchange, geometric response and initial constraints. Forming the real field before squaring retains interference between the carrier and addition. The cross contribution can reinforce or oppose the carrier's scalar response. Its sign and magnitude depend on the relative phase that propagation brings into the interaction region.

The scalar responds to successive electric and magnetic forces, so the observed change records their accumulated effect. Equal instantaneous sources can follow different histories and produce different scalar velocities. Geometry contributes through the changing normalization in Eq.~\eqref{eq:response}. Once the incident profiles and background are specified, the forced equations determine the response at each position and time, including its dependence on the mixing coefficient. Nonlinear evolution then tests this predicted response.

At quadratic order the energy exchanged into the scalar sector is
\begin{equation}
 \mathcal Q_{\psi A}^{(2)}
 =4\psi_0\psi_{0,T}(e^2-m^2).
 \label{eq:exchange}
\end{equation}
It appears with the opposite sign in the transverse Maxwell balance. Area, orbit shape and the background electric field supply the remaining work terms. The growing response draws energy from this coupled evolution. A component with very small initial self-energy can therefore affect a larger subsequent exchange. Its self-energy is one diagonal part of the initial energy; the interference with the carrier contributes separately to both the energy and the signed response.

\emph{Equal-energy selection.}
To isolate the role of coherence, we vary the low-frequency content of a compact packet while holding its initial electromagnetic energy fixed. The envelope is $f(u)=\exp[1-(1-u^2)^{-1}]$ for $|u|<1$, zero otherwise. The carrier's complex field and momentum come from black-hole transmission at $\omega M=0.2$. Its polarization maximizes magnetic transfer at $\delta=10^{-6}$ and is then held fixed. The added component has wavenumber $k/100$ and initial electric field $e_\ell=-\partial_yU_\ell$. Both components have the common amplitude $\delta^{1/2+\kappa}$ with $\kappa=1/2$, and $\tau_0=1$. The additions are prescribed local constrained data. Their profiles and phase conventions are given in the Supplemental Material~\cite{Supplement}.

Initially $r=0$ and the orbit-shape polarization vanishes. Each input is normalized to the same positive transverse Maxwell energy,
\begin{align}
 H_\perp&=2\int(e^2+U_y^2)\dd y,\notag\\
 N^2&=\frac{H_{cc}}{H_{cc}+\lambda H_{c\ell}+\lambda^2H_{\ell\ell}},
 \label{eq:energy}
\end{align}
where $H_{cc}$, $H_{c\ell}$ and $H_{\ell\ell}$ are the carrier, cross and low-frequency coefficients. Energy equality is imposed within each background, and the local constraints are solved for every input. Over an observation region $\mathcal O$, the response is the largest fractional change in normalized scalar velocity relative to the solution with no packet at the same time,
\begin{equation}
 A_\psi(T)=\max_{y\in\mathcal O}
 \frac{|p_\psi(T,y)-\pzero(T)|}{|\pzero(T)|}.
 \label{eq:amplitude}
\end{equation}

\begin{figure*}[t]
\includegraphics[width=\textwidth]{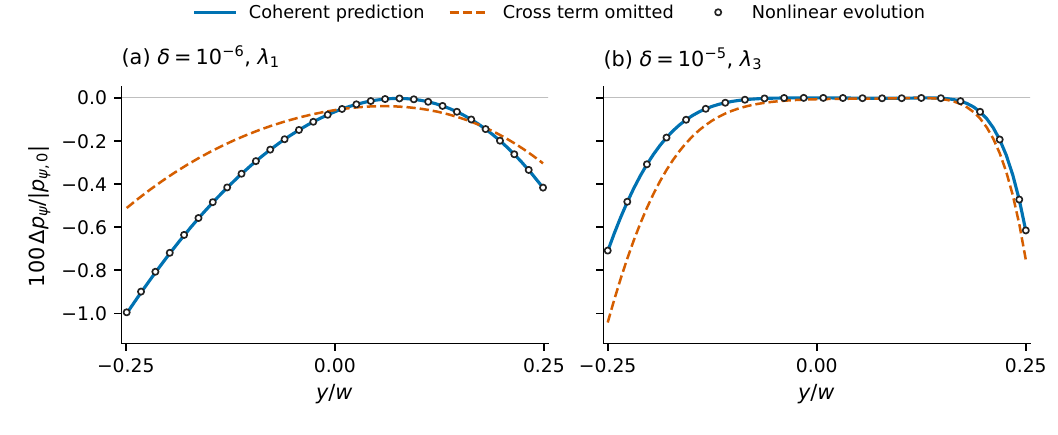}
\caption{Coherent enhancement and suppression of the signed scalar response. The cross contribution enhances the response in (a), $\delta=10^{-6}$, $\lambda_1$, $T\simeq242.4$, and partly cancels it in (b), $\delta=10^{-5}$, $\lambda_3$, $T\simeq53.39$. Solid and dashed curves show the quadratic prediction with and without cross terms; symbols show nonlinear evolution. Each panel uses common comparison points: 175 in $|y|\leq34.8$ and 65 in $|y|\leq w/4$, respectively. Symbols display a subset for visibility. Numerical uncertainty is smaller than the symbols.}
\label{fig:profiles}
\end{figure*}

On the reference background $\delta=10^{-6}$, increasing the low-frequency addition enhances the response. The mixing coefficients $\lambda_1\simeq8.103\times10^{-7}$ and $\lambda_3\simeq1.787\times10^{-5}$ were chosen from predicted responses of $1\%$ and $100\%$ at the predetermined time $T\simeq242.4$, over $|y|\leq34.8$. Nonlinear evolution gives about $0.995\%$ and $65.8\%$, preserving this ordering. For the smaller input, the coherent profile agrees to $0.52\%$, while omitting the cross term gives a $48.4\%$ discrepancy [Fig.~\ref{fig:profiles}(a)]. The larger input has entered the nonlinear regime beyond the quantitative second-order expansion.

At $\delta=10^{-5}$, the same increase in the admixture is predicted to suppress the response. Nonlinear evolution confirms this reversal at all three comparison times in Table~\ref{tab:ordering}. The ordering, times and 65 curves normal to the time slices in $|y|\leq w/4$ were fixed before nonlinear evolution. The first two times correspond to predicted $1\%$ and $3\%$ responses for $\lambda_1$; the last is the final common sampled prediction below $10\%$ within the prescribed observation window.

\begin{table}[t]
\caption{Relative response $100A_\psi$ for the two fixed inputs. P and N denote quadratic prediction and nonlinear evolution. The asterisk marks the strong reference prediction beyond its quantitative range. Displayed times are rounded; the fixed comparison surfaces are specified in the Supplemental Material. Uncertainty in response threshold times is set by temporal resolution.}
\label{tab:ordering}
\begin{ruledtabular}
\begin{tabular}{cccccc}
 & & \multicolumn{2}{c}{$\lambda_1$} & \multicolumn{2}{c}{$\lambda_3$}\\
$\delta$ & $T$ & P & N & P & N\\
$10^{-6}$ & 242.4 & 1.000 & 0.995 & $100.0^{*}$ & 65.77 \\
$10^{-5}$ & 53.39 & 1.000 & 0.995 & 0.711 & 0.708 \\
$10^{-5}$ & 54.06 & 3.000 & 2.953 & 2.135 & 2.111 \\
$10^{-5}$ & 54.75 & 9.556 & 9.097 & 6.806 & 6.569 \\
\end{tabular}
\end{ruledtabular}
\end{table}

For $\lambda_3$ on the changed background, the low-frequency self-energy is $4.70\times10^{-13}$ of the carrier energy. The initial cross-energy fraction is $-5.36\times10^{-11}$, and $N-1=2.66\times10^{-11}$. The geometric initial spectra of the two inputs agree to $2.3\times10^{-16}$. These small initial differences accompany roughly $29\%$ suppression of the first response relative to $\lambda_1$. Evaluating the accumulated response at $y=-w/4$ gives the carrier, cross and low-frequency contributions
\begin{align}
 100\frac{\Delta p_\psi^{(2)}}{|\pzero|}
 &\simeq-1.015+0.331-0.027\notag\\
 &=-0.711.
 \label{eq:cancellation}
\end{align}
The positive cross term opposes the negative carrier response [Fig.~\ref{fig:profiles}(b)]. This cancellation is measured on the fixed observation curve at the region's edge.

The prediction also captures the early magnitude and timing. Its profile discrepancies at the three times are $0.38\%$, $1.12\%$ and $3.48\%$, compared with $46.9\%$, $47.5\%$ and $49.8\%$ after omitting cross terms. Numerical uncertainty is below $1.66\times10^{-7}$ in absolute $p_\psi$, resolving the growing discrepancy as the amplitude expansion breaks down. The predicted separation between the two inputs' $1\%$ and $3\%$ response times is about $0.2$ in $T$. Nonlinear interpolation preserves it at a sampling interval of at most $0.0742$.

The reversal follows from the combined change in geometry, envelope and transmitted phase. Here $kw$ decreases from $200$ to about $63.25$, while $w$ decreases from about $139.8$ to $47.54$. The ratio of the instantaneous pump to $k^2$ at the comparison endpoints changes from approximately $0.0257$ to $0.427$. These changes alter the field's history and the sign of its accumulated cross contribution. A vacuum Kasner approximation reproduces the charged-background prediction in the latter window to about $1.15\times10^{-8}$ relatively, so residual charge supplies a small correction there.

The fixed mixing coefficients, polarization and physical packet rule make the change of background a predictive test. Its altered transmission and propagation determine the new bilinear coefficients before the nonlinear response is known. The initial size of the added component alone would suggest the wrong ordering on the second background; the coherent calculation follows the sign of the accumulated contribution. In this comparison width and transmission phase vary together. Their combined effect selects opposite responses, while varying them independently would distinguish their individual roles.

\emph{First nonlinear exchange.}
The selected response evolves into a first magnetic pulse. For the two reference inputs, its central maxima lie near logarithmic volume times $s=285.1$ and $269.9$, where $s=-\Delta(r+\sigma)$ along each normal curve. The magnetic field slows the scalar, after which the electric field grows again. Integrated electric and magnetic contributions have opposite signs, and longitudinal terms remain appreciable. Through $T=380$, the monitored curves show a resolved exchange without an outgoing Kasner plateau meeting the criteria in the Supplemental Material~\cite{Supplement}.

An isolated homogeneous magnetic layer gives an analytic reference for this interaction. With a self-consistent single magnetic axis, the scalar reverses while the geometric rates receive a small impulse. An exact scalar-energy identity bounds that impulse and gives a tangent correction of order $q_B^{-1}$, where $q_B$ is half the scalar value $\varphi=\sqrt2\psi$ at the actual magnetic maximum. The interval conditions, accumulated remainders and eight controls with two Kasner plateaus are given in the Supplemental Material~\cite{Supplement}. Propagation selects the response preceding the magnetic layer; continuing electric and spatial exchange determines whether its later evolution approaches this isolated reflection.

\emph{Conclusion.}
The first response of this critical Kasner family remembers how the incident electromagnetic field was prepared. Propagation builds a signed interference contribution, allowing a weak spectral addition to enhance or suppress the scalar response. Predicting the reversal between backgrounds connects incoming data to nonlinear geometry without fitting the evolved response. The critical scale gives this memory a physical setting: longitudinal transport remains effective during the estimated magnetic delay, and the packet's prescribed duration fixes how many carrier oscillations it contains. This connection between incident coherence and geometric response extends the questions addressed by stationary critical interiors and controlled inhomogeneous bounce models~\cite{LiSunYang2026,Li2024Surface}.

Following the mechanism from a black-hole horizon requires the regular incoming radiation to retain the relevant low-frequency field and momentum projection. The induced metric and scalar fields must accompany that radiation through the matching region. A second question concerns the first exchange itself: sufficient isolation would produce a second Kasner plateau and allow comparison with the magnetic reflection law. These extensions connect the predicted response to the radiation that reaches the interior and to the matter exchange that follows~\cite{Burko2003,BurkoErratum2003,Henneaux2022}.

\paragraph*{Data availability.}
The data and numerical source code supporting this work are available at \url{https://github.com/l46430640-del/public_archive_for_Kasner}.

\paragraph*{Use of AI tools.}
OpenAI Codex assisted with numerical code generation and language editing. The author has checked these contributions and takes full responsibility for this work.
\onecolumngrid
\section*{End Matter}
\label{sec:endmatter}
\twocolumngrid
\subsection*{Quadratic Response and Initial Data}
With a conformal base, the polarized orbit metric is $\rho\,\diag(e^{P_{\rm o}},e^{-P_{\rm o}})$ with area $\rho=e^r$. The dynamical orbit connection has zero conserved orbit momenta, and the electric momentum is $q$. The full metric, action and constraints appear in the Supplemental Material~\cite{Supplement}.

The homogeneous zero-packet fields satisfy
\begin{align}
 r_{0,TT}&=b-r_{0,T}^2,\notag\\
 \sigma_{0,TT}&=-b+r_{0,T}^2/4-\psi_{0,T}^2,\notag\\
 \psi_{0,TT}&=\psi_0b/2-r_{0,T}\psi_{0,T},
 \label{eq:embackground}
\end{align}
with $b=(q^2/8)e^{-\psi_0^2-2r_0+2\sigma_0}$. Initially $r_0=\sigma_0=0$, $\psi_0=\psi_K$, $r_{0,T}=-2p_s$, $\psi_{0,T}=v$. The constraint fixes $\sigma_{0,T}=(b-r_{0,T}^2/2+2v^2)/(2r_{0,T})$. The parameter triples $(\beta,\psi_K,q)$ are approximately $(76.84,0.3944,1.464\times10^{-6})$ and $(24.37,1.051,1.557\times10^{-5})$.

Let $R,\Sigma,D,\Phi$ be the second-order changes of $r,\sigma,P_{\rm o},\psi$, and put $\delta b=b(-2\psi_0\Phi-2R+2\Sigma)$. The fields $U$ and $U_T$ propagated by Eq.~\eqref{eq:linear} source the response equations
\begin{align}
 R_{TT}-R_{yy}&=-2r_{0,T}R_T+\delta b+4be^{-r_0}U^2,\notag\displaybreak[1]\\
 \Sigma_{TT}-\Sigma_{yy}&=\tfrac12r_{0,T}R_T-2\psi_{0,T}\Phi_T\notag\\
 &\quad-\delta b-6be^{-r_0}U^2,\notag\displaybreak[1]\\
 D_{TT}-D_{yy}&=-r_{0,T}D_T-2e^{-r_0}(e^2-m^2)\notag\\
 &\quad+4be^{-r_0}U^2,\notag\displaybreak[1]\\
 \Phi_{TT}-\Phi_{yy}&=-r_{0,T}\Phi_T-\psi_{0,T}R_T
 +\tfrac12b\Phi\notag\\
 &\quad+\tfrac12\psi_0\delta b+\psi_0e^{-r_0}(e^2-m^2).
 \label{eq:emforced}
\end{align}
Initially $R=D=\Phi=R_T=D_T=\Phi_T=0$. The two local metric constraints give
\begin{align}
 \Sigma_y&=\frac{2e^{-r_0}em}{r_{0,T}},\qquad \Sigma(T=0,y=0)=0,\notag\\
 \Sigma_T&=\frac{b\Sigma+2be^{-r_0}U^2+e^{-r_0}(e^2+m^2)}{r_{0,T}}.
 \label{eq:eminitial}
\end{align}
For the nonlinear initial data, the corresponding relations are $\sigma_y=2em/r_T$ and $\sigma_T=[b+4bU^2-r_T^2/2+2v^2+2(e^2+m^2)]/(2r_T)$ at $r=P_{\rm o}=0$, with $\sigma(T=0,y=0)=0$ and $b=q^2e^{-\psi_K^2+2\sigma}/8$ evaluated locally.

The compact profiles retain the complex transferred field $B$ and momentum coefficient $E$ of the $\ell=2$ odd carrier. Set $c=B/(-ik)$, $A_0=\delta^{1/2+\kappa}$ and
\begin{align}
 U_c&=A_0\Re[ce^{i\vartheta-iky}]f(y/w),\notag\\
 e_c&=-A_0\Re[Ee^{i\vartheta-iky}]f(y/w),\notag\\
 U_\ell&=A_0|c|\Re[e^{i\vartheta-iky/100}]f(y/w),\qquad e_\ell=-U_{\ell,y}.
 \label{eq:empacket}
\end{align}
The main comparison uses $\vartheta=0$. Initially $U=N(U_c+\lambda U_\ell)$, $e=N(e_c+\lambda e_\ell)$, $m=U_y$, and $U_T=e+gU$. In particular $m$ includes the envelope derivative. The carrier coefficients are tabulated in the Supplemental Material~\cite{Supplement}.

The energy coefficients are $H_{cc}=2\int(e_c^2+m_c^2)\dd y$, $H_{c\ell}=4\int(e_ce_\ell+m_cm_\ell)\dd y$, and $H_{\ell\ell}=2\int(e_\ell^2+m_\ell^2)\dd y$. The self-energy fraction is evaluated as $\lambda^2H_{\ell\ell}/H_{cc}$. In Eq.~\eqref{eq:quadratic}, $C_{c\ell}$ includes the factor of two from each cross source and its initial constraint. Setting this coefficient to zero defines the competing prediction for the same full physical field.

\subsection*{Accuracy of the Decisive Comparison}
The nonlinear equations use centered fourth-order spatial differences and fourth-order Runge--Kutta time integration. For $\delta=10^{-5}$, three grids have spacings $w/128,w/256,w/512$, and independent time steps at $h=w/256$ have $\Delta T/h=0.4,0.2,0.1$. Spatial convergence is approximately fourth order, with middle-to-fine spectrum differences below $1.93\times10^{-9}$. The finest independent time-step difference is below $3.82\times10^{-11}$. Normalized constraint residuals are below $6.14\times10^{-11}$. The domains cover the causal past of the observation curves; doubling the domain leaves their spectra unchanged in double precision.

Predictions and nonlinear responses use common fixed times and curves. For $\lambda_3$ at the three times on the $\delta=10^{-5}$ background, combined absolute numerical estimates in $p_\psi$ are $1.44\times10^{-8}$, $1.64\times10^{-8}$ and $1.34\times10^{-7}$. They include refinement and temporal interpolation errors, and are much smaller than the second-order expansion discrepancy. The continuous initial energy is normalized by quadrature; its finest finite-difference error is approximately $7.86\times10^{-6}$ relatively and is shared by the two inputs to much greater precision than the measured difference.

Sixth-order differences with DOP853 evolve the $\delta=10^{-5}$ comparison independently from its initial surface. Its full-spectrum discrepancy is below $1.29\times10^{-10}$. An independent Maxwell calculation in continuous frequency retains complex fields and momenta; refining frequency spacing, cutoff and initial quadrature changes the combined field by less than $6.28\times10^{-10}$ relatively. Comparison to the spatial discretization differs by less than $6.3\times10^{-8}$.

The reconstructed four-dimensional fields give relative Einstein and Maxwell residuals below $9.21\times10^{-10}$ and $1.43\times10^{-10}$, and a scalar residual below $8.86\times10^{-13}$ after division by squared expansion. The reduced-energy balance residual is below $9.35\times10^{-12}$. These derivative-fit checks are not uniformly monotone under refinement~\cite{Supplement}; the output error estimate comes from the independent evolution and resolution comparisons above.

\onecolumngrid
\clearpage
\setcounter{equation}{0}
\setcounter{table}{0}
\setcounter{figure}{0}
\setcounter{section}{0}
\setcounter{subsection}{0}
\setcounter{secnumdepth}{3}
\renewcommand{\theequation}{S\arabic{equation}}
\renewcommand{\thetable}{S\Roman{table}}
\renewcommand{\thefigure}{S\arabic{figure}}
\renewcommand{\theHequation}{supplement.\arabic{equation}}
\renewcommand{\theHtable}{supplement.\arabic{table}}
\renewcommand{\theHfigure}{supplement.\arabic{figure}}
\renewcommand{\theHsection}{supplement.\arabic{section}}
\begin{center}
\large\bfseries Supplemental Material for\\
Coherent Selection of Critical Kasner Response
\end{center}
\twocolumngrid

\section{Conventions and Response Observables}

The local EMS model has two commuting spacelike Killing fields and supports propagation in the remaining spatial direction. We derive its constrained equations, construct the coherent response and give the numerical comparisons below. The stationary critical black-hole family supplies background parameters~\cite{sm:LiSunYang2026}; the nonlinear experiments use wave packets satisfying the local constraints.

In signature $(-,+,+,+)$ and $G=c=1$, the action is
\begin{equation}
 I=\frac{1}{16\pi}\int\dd^4x\sqrt{-g}
 [R-2(\nabla\psi)^2-ZF_{\mu\nu}F^{\mu\nu}],\quad Z=e^{\psi^2}.
 \label{sm:eq:action}
\end{equation}
The metric and Maxwell potential take the form
\begin{align}
 \dd s^2&=e^{2\sigma}(-\dd T^2+\dd y^2)+\rho S_{AB}\theta^A\theta^B,\notag\\
 \theta^A&=\dd x^A+C^A_a\dd x^a,\qquad \det S=1,\notag\\
 A&=b_a\dd x^a+a_A\theta^A,\qquad r=\ln\rho.
 \label{sm:eq:metric}
\end{align}
The Killing indices are $A,B=1,2$, while $a,b$ are base indices. We work in the sector $F_{AB}=0$. The orbit connection remains dynamical. The geometric spectrum on the contracting foliation is
\begin{equation}
 P=\left(\operatorname{eig}\frac{K^i{}_j}{\tr K};p_\psi\right),
 \qquad p_\psi=\frac{\sqrt2\,n^\mu\nabla_\mu\psi}{|\tr K|}.
 \label{sm:eq:spectrum}
\end{equation}
Here $n=e^{-\sigma}\partial_T$ points toward contraction, $K_{ij}=\mathcal L_n\gamma_{ij}/2$, and $\tr K=e^{-\sigma}(r_T+\sigma_T)<0$ on the reference solution. The canonical scalar is $\varphi=\sqrt2\psi$. A free epoch obeys $\sum p_a=1$ and $\sum p_a^2+p_\psi^2=1$.

The stationary background family has $\delta=(Q_{\rm BH}/M)/q_c-1$, $q_c\simeq0.99396434$, and $\beta\sqrt\delta\to c_0\simeq0.076817$. Its exponents are $p_s=2/(\beta^2+3)$, $p_\parallel=(\beta^2-1)/(\beta^2+3)$ and $p_\psi=2\sqrt2\beta/(\beta^2+3)$; write $v=p_\psi/\sqrt2$. A free-flight magnetic proxy is defined by $L_B=\ln|\mathcal A_B|^{-1}$ and
\begin{align}
 \ln\mathcal R_B(t)&=-2L_B-2p_st+(\psi_K+vt)^2-\psi_K^2,\notag\\
 t_*&=\frac{p_s-\psi_Kv+\sqrt{(p_s-\psi_Kv)^2+2v^2L_B}}{v^2}.
 \label{sm:eq:tstar}
\end{align}
The second line solves $\mathcal R_B=1$. It is a free-flight threshold, distinct from the nonlinear magnetic maximum. For the prescribed physical packet rule, $k=\omega/\Lambda$, $w=\Lambda M\tau_0/\sqrt\delta$ and $kw=(\omega M)\tau_0/\sqrt\delta$, where $\Lambda=a_\parallel|\tr K|$ initially.

For the charged zero-packet background let $U=e^{\psi_0^2/2}a_1$, $g=\psi_0\psi_{0,T}$ and $b=(q^2/8)e^{-\psi_0^2-2r_0+2\sigma_0}$. The propagated fields obey
\begin{align}
 U_{TT}-U_{yy}&=(g_T+g^2-2b)U,\notag\\
 e&=U_T-gU,\qquad m=U_y.
 \label{sm:eq:linear}
\end{align}
Let $R,\Sigma,D,\Phi$ denote second-order changes of $r,\sigma$, orbit shape and $\psi$. The scalar response is
\begin{equation}
 \Delta p_\psi^{(2)}=\frac{\sqrt2}{|\calK_0|}
 \left[\Phi_T-\frac{\psi_{0,T}}{\calK_0}(R_T+\Sigma_T)\right],
 \label{sm:eq:predictor}
\end{equation}
Here $\calK_0=r_{0,T}+\sigma_{0,T}<0$.
For $U=N(U_c+\lambda U_\ell)$, its bilinear expansion is
\begin{equation}
 \Delta p_\psi^{(2)}=N^2(C_{cc}+\lambda C_{c\ell}+\lambda^2C_{\ell\ell}).
 \label{sm:eq:quadratic}
\end{equation}
All three coefficients include geometry and initial constraints. The cross coefficient includes twice the product of the two fields. Equal energy is imposed initially at $r=P_{\rm o}=0$:
\begin{align}
 H_\perp&=2\int(e^2+U_y^2)\dd y,\notag\\
 N^2&=\frac{H_{cc}}{H_{cc}+\lambda H_{c\ell}+\lambda^2H_{\ell\ell}}.
 \label{sm:eq:energyinput}
\end{align}
In the complete reduction below, $P_{\rm o}$ is denoted simply by $P$ as an orbit-shape coordinate; the geometric spectrum is the four-vector defined in Eq.~\eqref{sm:eq:spectrum}.

The response comparison uses $\Delta p_\psi=p_\psi-\pzero$ at the same conformal time on the zero-packet solution and $A_\psi=\max_{\mathcal O}|\Delta p_\psi|/|\pzero|$. The profile discrepancy is $\max_{\mathcal O}|\Delta p_\psi^{(2)}-\Delta p_\psi|/\max_{\mathcal O}|\Delta p_\psi^{(2)}|$. The omitted-cross discrepancy uses the same denominator and the same observation points.

\begin{table*}[t]
\caption{Initial parameters for Eq.~\eqref{sm:eq:packetexplicit} and the charged background. $\psi_K$ is the initial scalar and $q$ the reduced electric momentum. $B$ and $E$ include the carrier's complex transfer phase. Extra digits specify the input.}
\label{sm:tab:initialparams}
\begin{ruledtabular}
\begin{tabular}{crr}
 & $\delta=10^{-6}$ & $\delta=10^{-5}$\\
$\beta$ & $76.8408656$ & $24.3664773$ \\
$\psi_K$ & $0.394434995$ & $1.05144053$ \\
$q$ & $1.46375513\times10^{-6}$ & $1.55738295\times10^{-5}$ \\
$k$ & $1.43080143$ & $1.3304029$ \\
$w$ & $139.7818$ & $47.5386466$ \\
$\Re B$ & $1.65943217$ & $-1.07527339$ \\
$\Im B$ & $1.10947142\times10^{-16}$ & $-1.16997205$ \\
$\Re E$ & $-1.63478468$ & $1.1286426$ \\
$\Im E$ & $0.0143399857$ & $1.1122244$ \\
\end{tabular}
\end{ruledtabular}
\end{table*}
\section{Reduction and Local Constraints}
\label{sm:app:reduction}

\subsection{Variation before gauge fixing}

The four-dimensional field equations following from Eq.~\eqref{sm:eq:action} are
\begin{align}
 G_{\mu\nu}&=2\psi_{,\mu}\psi_{,\nu}-g_{\mu\nu}(\nabla\psi)^2\notag\\
 &\quad+2Z\left(F_{\mu\rho}F_\nu{}^\rho-\tfrac14g_{\mu\nu}F^2\right),\notag\\
 \nabla_\mu(ZF^{\mu\nu})&=0,\qquad
 4\Box\psi-Z'F^2=0.
 \label{sm:eq:fourdim}
\end{align}
Before fixing the base metric, write $G_{AB}=\rho S_{AB}$ and $H^A=\dd C^A$. In the sector $F_{AB}=0$ the field strength decomposes as
\begin{equation}
 F=\mathcal F+\dd a_A\wedge\theta^A,
 \qquad \mathcal F=\dd b+a_AH^A.
 \label{sm:eq:decomposition}
\end{equation}
Integration over a unit coordinate area of the Killing orbits gives, up to boundary terms and the overall $1/(16\pi)$,
\begin{align}
 \frac{\mathcal L_2}{\sqrt{-h}}={}&\rho R[h]+\frac{(\nabla\rho)^2}{2\rho}
 \notag\\
 &-\frac{\rho}{4}\tr(S^{-1}\nabla S S^{-1}\nabla S)
 -2\rho(\nabla\psi)^2\notag\\
 &-2ZS^{AB}\nabla a_A\nabla a_B
 \notag\\
 &-\frac{\rho G_{AB}}4H^A_{ab}H^{B\,ab}
 -\rho Z\mathcal F_{ab}\mathcal F^{ab}.
 \label{sm:eq:reducedaction}
\end{align}
All contractions in this expression use $h_{ab}$. Variation of its lapse and shift gives the Hamiltonian and momentum constraints, including the Maxwell momentum density. The trace-free base equations give their null form below.

Choose $h_{ab}=e^{2\sigma}\diag(-1,1)$, $C^A_T=b_T=0$ and
\begin{equation}
 S=\begin{pmatrix}e^P&e^PQ\\e^PQ&e^{-P}+e^PQ^2\end{pmatrix}.
 \label{sm:eq:shape}
\end{equation}
The electric momentum $q$ and orbit momenta $J_A$ are constant on the base. A Routh transformation, which fixes these momenta before eliminating their velocities, yields
\begin{align}
 \mathcal L&=\tfrac12X_T^{\mathsf t}\mathsf K X_T
            -\tfrac12X_y^{\mathsf t}\mathsf K X_y-V,
 \notag\\
 X&=(r,\sigma,P,Q,\psi,a_1,a_2).
 \label{sm:eq:routh}
\end{align}
The nonzero entries of the symmetric matrix $\mathsf K$ are
\begin{align}
 \mathsf K_{rr}&=-\rho,&\mathsf K_{r\sigma}&=-2\rho,\notag\\
 \mathsf K_{PP}&=\rho,&\mathsf K_{QQ}&=\rho e^{2P},\notag\\
 \mathsf K_{\psi\psi}&=4\rho,&
 \mathsf K_{a_Aa_B}&=4ZS^{AB},
 \label{sm:eq:kinetic}
\end{align}
and
\begin{equation}
 V=e^{2\sigma}\left[\frac{q^2}{8\rho Z}
 +\frac{(J-qa)^{\mathsf t}S^{-1}(J-qa)}{2\rho^2}\right].
 \label{sm:eq:potential}
\end{equation}
The sign and $a$ dependence of the second term are needed even when $J=0$. The reconstructed potentials satisfy
\begin{align}
 \partial_TC_y&=e^{2\sigma}\rho^{-2}S^{-1}(J-qa),\notag\\
 \partial_Tb_y&=\frac{q e^{2\sigma}}{4\rho Z}-a^{\mathsf t}\partial_TC_y.
 \label{sm:eq:reconstruct}
\end{align}
Initial constants in these auxiliary potentials fix coordinate and Maxwell gauge choices.

\subsection{Nonlinear wave equations}

Writing $v^i=X_T^i$ and $w^i=X_y^i$, the seven nonlinear equations follow from the kinetic matrix and potential in Eqs.~\eqref{sm:eq:kinetic}--\eqref{sm:eq:potential}:
\begin{align}
 \mathsf K_{ij}(X_{TT}^j-X_{yy}^j)
 ={}&\tfrac12\mathsf K_{jk,i}(v^jv^k-w^jw^k)\notag\\
 &-\mathsf K_{ij,k}(v^kv^j-w^kw^j)-V_{,i}.
 \label{sm:eq:completewaves}
\end{align}
This expression includes all derivatives of the field-dependent kinetic matrix. Together with the reconstruction equations it determines the local metric and electromagnetic field.

For $D_\pm=\partial_T\pm\partial_y$, the two constraints read
\begin{align}
 0=\frac{C_\pm}{\rho}={}&D_\pm^2r+\tfrac12(D_\pm r)^2
 -2D_\pm\sigma D_\pm r\notag\\
 &+\tfrac12[(D_\pm P)^2+e^{2P}(D_\pm Q)^2]
 +2(D_\pm\psi)^2\notag\\
 &+2Z\rho^{-1}(D_\pm a)^{\mathsf t}S^{-1}(D_\pm a).
 \label{sm:eq:constraints}
\end{align}
Second time derivatives in an initial constraint are eliminated with Eq.~\eqref{sm:eq:completewaves}. The two constraints and the residual base coordinate freedom remove four phase-space functions from the fourteen in $X,X_T$. The remaining five wave degrees of freedom comprise two gravitational, two electromagnetic and one scalar mode. The conserved momenta label sectors rather than additional local waves.

The numerical experiments use the invariant subspace $Q=a_2=0$, $J_A=0$. Initially $r=P=0$, $P_T=0$, $\psi=\psi_K$, $r_T=-2p_s$ and $\psi_T=v$, with no spatial variation of these fields. The freely specified field and electric amplitude are $U$ and $e$, where $a_1=e^{-\psi_K^2/2}U$ and $a_{1,T}=e^{-\psi_K^2/2}e$. The momentum constraint fixes
\begin{equation}
 \sigma_y=\frac{2e\,U_y}{r_T},\qquad \sigma(0)=0.
 \label{sm:eq:initialsigma}
\end{equation}
Here $\sigma(0)$ denotes the center $y=0$ on the initial surface. Define $b=q^2e^{-\psi_K^2+2\sigma}/8$ on this surface. The average null constraint then fixes
\begin{equation}
 \sigma_T=\frac{b+4bU^2-\tfrac12r_T^2+2v^2+2(e^2+U_y^2)}{2r_T}.
 \label{sm:eq:initialvelocity}
\end{equation}
These local data include the metric response induced by the packet energy and flux. No additional independent gravitational or scalar pulse is imposed.

\section{Linear and Quadratic Response}
\label{sm:app:response}

\subsection{The charged background}

The homogeneous zero-packet solution has $P=Q=a_A=J_A=0$. Its equations are
\begin{align}
 r_{0,TT}&=b-r_{0,T}^2,\notag\\
 \sigma_{0,TT}&=-b+r_{0,T}^2/4-\psi_{0,T}^2,\notag\\
 \psi_{0,TT}&=\psi_0b/2-r_{0,T}\psi_{0,T}.
 \label{sm:eq:background}
\end{align}
We initialize $r_0=\sigma_0=0$, $\psi_0=\psi_K$, $r_{0,T}=-2p_s$, $\psi_{0,T}=v$. Equation~\eqref{sm:eq:initialvelocity} with $U=e=0$ fixes $\sigma_{0,T}$. Thus the charged background itself satisfies the constraint.

Linearizing the Maxwell sector gives
\begin{equation}
 a_{1,TT}-a_{1,yy}+2g a_{1,T}+2b a_1=0.
 \label{sm:eq:linearoriginal}
\end{equation}
Substituting $a_1=U/\sqrt{Z_0}$ gives Eq.~\eqref{sm:eq:linear}. In Fourier space both $\widetilde U(T,k)$ and $\partial_T\widetilde U(T,k)$ evolve. Real initial data require the negative-frequency conjugates; forming the real sum before squaring retains sum-frequency and difference-frequency sources.

\subsection{Second-order geometry and scalar}

Set $r=r_0+R$, $\sigma=\sigma_0+\Sigma$, $P=D$ and $\psi=\psi_0+\Phi$ to second order in the packet amplitude. Write
\begin{equation}
 \delta b=b(-2\psi_0\Phi-2R+2\Sigma),\qquad
 \calS=e^2-m^2.
 \label{sm:eq:deltab}
\end{equation}
The full forced system at this order is
\begin{align}
 R_{TT}-R_{yy}&=-2r_{0,T}R_T+\delta b+4be^{-r_0}U^2,\notag\\
 \Sigma_{TT}-\Sigma_{yy}&=\tfrac12r_{0,T}R_T-2\psi_{0,T}\Phi_T\notag\\
 &\quad-\delta b-6be^{-r_0}U^2,\notag\\
 D_{TT}-D_{yy}&=-r_{0,T}D_T-2e^{-r_0}\calS+4be^{-r_0}U^2,\notag\\
 \Phi_{TT}-\Phi_{yy}&=-r_{0,T}\Phi_T-\psi_{0,T}R_T\notag\\
 &\quad+\tfrac12b\Phi+\tfrac12\psi_0\delta b
 +\psi_0e^{-r_0}\calS.
 \label{sm:eq:secondorder}
\end{align}
The initial values of $R,D,\Phi$ and their velocities vanish. At the initial surface, expanding Eqs.~\eqref{sm:eq:initialsigma} and \eqref{sm:eq:initialvelocity} gives
\begin{align}
 \Sigma_y&=\frac{2e^{-r_0}em}{r_{0,T}},\qquad\Sigma(0)=0,\notag\\
 \Sigma_T&=\frac{b\Sigma+2be^{-r_0}U^2+e^{-r_0}(e^2+m^2)}{r_{0,T}}.
 \label{sm:eq:secondinitial}
\end{align}
Expanding Eq.~\eqref{sm:eq:spectrum} on the contracting branch yields Eq.~\eqref{sm:eq:predictor}. This expansion includes the metric response already present on the initial surface. The zero-packet comparator used in all response differences is the solution of Eq.~\eqref{sm:eq:background} at the same $T$.

For each input pair, substituting $(U_c,e_c,m_c)$, the cross products, and $(U_\ell,e_\ell,m_\ell)$ into Eqs.~\eqref{sm:eq:secondorder} and \eqref{sm:eq:secondinitial} gives three forced problems. Their linearity yields the coefficients in Eq.~\eqref{sm:eq:quadratic}. Setting the cross coefficient to zero defines a competing prediction for the same initial data and Maxwell evolution containing both components.

\subsection{Packet specification}

Let $B$ and $E$ be the complex carrier coefficients projected into the compatible local polarization, in initial expansion units, and let $c=B/(-ik)$. The envelope is $f(u)=\exp[1-(1-u^2)^{-1}]$ for $|u|<1$, and zero otherwise. With $A_0=\delta^{1/2+\kappa}$ and phase $\vartheta$, the initial data are
\begin{align}
 U_c&=A_0\Re[ce^{i\vartheta-iky}]f(y/w),\notag\\
 e_c&=-A_0\Re[Ee^{i\vartheta-iky}]f(y/w),\notag\\
 U_\ell&=A_0|c|\Re[e^{i\vartheta-iky/100}]f(y/w),\notag\\
 e_\ell&=-\partial_yU_\ell,\qquad
 (U,e)=N(U_c+\lambda U_\ell,e_c+\lambda e_\ell).
 \label{sm:eq:packetexplicit}
\end{align}
In particular $m_c=\partial_yU_c$ includes the envelope derivative. The initial linear velocity is $U_T=e+gU$, whereas the nonlinear potential velocity is $a_{1,T}=e/\sqrt Z$ on this initial surface. These two prescriptions are equivalent in their respective variables.

The coefficients of Eq.~\eqref{sm:eq:energyinput} are
\begin{align}
 H_{cc}&=2\int(e_c^2+m_c^2)\dd y,\notag\\
 H_{c\ell}&=4\int(e_ce_\ell+m_cm_\ell)\dd y,\notag\\
 H_{\ell\ell}&=2\int(e_\ell^2+m_\ell^2)\dd y.
 \label{sm:eq:energycoefs}
\end{align}
The fraction $\lambda^2H_{\ell\ell}/H_{cc}$ is evaluated directly to resolve the tiny low-frequency self-energy accurately. The Maxwell evolution retains the full frequency content of the finite envelope.

Table~\ref{sm:tab:initialparams} gives the dimensionless background and carrier parameters. The main comparisons use $\vartheta=0$. Changing the reference input $\lambda_3$ to $\vartheta=\pi/2$ gives a $73.5\%$ nonlinear response at $T=242.4484$, compared with $65.8\%$ at phase zero. Its quadratic estimate is about $119\%$, beyond the quantitative expansion. This comparison shows phase sensitivity of the nonlinear event; the early reversal test uses phase zero.

\section{Energy Exchange and the Geometric Response}
\label{sm:app:exchange}

The autonomous reduced action has the local Noether balance
\begin{align}
 \partial_T\calH-\partial_y\mathcal J&=0,\notag\\
 \calH&=\tfrac12v^{\mathsf t}\mathsf K v+\tfrac12w^{\mathsf t}\mathsf K w+V,\notag\\
 \mathcal J&=v^{\mathsf t}\mathsf K w.
 \label{sm:eq:noether}
\end{align}
The full gravitational kinetic form is indefinite. The positive transverse electromagnetic density is instead
\begin{equation}
 h_A=2Z(a_T^{\mathsf t}S^{-1}a_T+a_y^{\mathsf t}S^{-1}a_y),
 \label{sm:eq:emenergy}
\end{equation}
which reduces to the integrand of $H_\perp$ initially. The scalar density is $h_\psi=2\rho(\psi_T^2+\psi_y^2)$. Multiplying the scalar and Maxwell equations by their velocities gives opposite exchange terms
\begin{equation}
 \mathcal Q_{\psi A}=2Z'\psi_T
 (a_T^{\mathsf t}S^{-1}a_T-a_y^{\mathsf t}S^{-1}a_y).
 \label{sm:eq:exchange}
\end{equation}
It enters the scalar balance with a plus sign and the transverse Maxwell balance with a minus sign. Area and shape evolution, the electric and orbit potential, and spatial flux supply the remaining work terms. Equation~\eqref{sm:eq:noether} tests their sum. Neither this reduced balance nor $H_\perp$ is an ADM mass balance.

For the first-event source decomposition, define $\calK=r_T+\sigma_T$ and the dimensionless field fractions
\begin{equation}
 \calE=\frac{Ze^{-r-P}a_{1,T}^2}{\calK^2},\qquad
 \calB=\frac{Ze^{-r-P}a_{1,y}^2}{\calK^2}.
 \label{sm:eq:fractions}
\end{equation}
They are the transverse electromagnetic densities divided by the squared expansion in this normalization. Their direct contributions to the normalized scalar rate are
\begin{equation}
 \dot p_{\psi,E}=\sqrt2\psi|\calK|\calE,
 \qquad \dot p_{\psi,B}=-\sqrt2\psi|\calK|\calB,
 \label{sm:eq:emrates}
\end{equation}
where the dot here denotes $\partial_T$. For any acceleration vector $A^i=X_{TT}^i$, the scalar component of the geometric rate is
\begin{equation}
 \mathcal D_\psi[v;A]=\frac{\sqrt2}{|\calK|}
 \left[A^\psi-v^\psi\frac{A^r+A^\sigma}{\calK}\right].
 \label{sm:eq:ratefunctional}
\end{equation}
This map is linear in $A$ at fixed state.

Let $F(X,v,w;q,J)$ be the nonlinear force $X_{TT}-X_{yy}$ in Eq.~\eqref{sm:eq:completewaves}, and let $w_m$ retain only the Maxwell entries of $w$. The acceleration splits into
\begin{align}
 A_0&=F(X,v_0,0;0,0),\notag\\
 A_m&=F(X,v,w_m;q,J)-A_0,\notag\\
 A_\nabla&=X_{yy}+F(X,v,w;q,J)-F(X,v,w_m;q,J),
 \label{sm:eq:partition}
\end{align}
where $v_0$ is $v$ with its Maxwell entries set to zero. The rates are $\mathcal D_\psi[v;A_0]$, $\mathcal D_\psi[v;A_m]$ and $\mathcal D_\psi[v;A_\nabla]$. The magnetic first derivative stays in $A_m$ as a physical field, while the remaining scalar and geometric gradients enter $A_\nabla$. Subtracting Eq.~\eqref{sm:eq:emrates} from the matter rate leaves the small charge and orbit contribution. Integrating these rates in $T$ over the stated intervals gives Table~\ref{sm:tab:pulse}.

\begin{table}[t]
\caption{Vacuum geometric and remaining matter contributions to the first exchange. $I_0$ and $I_q$ denote their respective integrals. The quadrature residual $R_I=I_E+I_B+I_\nabla+I_0+I_q-\Delta p_\psi$ measures integration over sampled times; evolution errors are estimated separately.}
\label{sm:tab:closure}
\begin{ruledtabular}
\begin{tabular}{crrr}
Input & $I_0$ & $I_q$ & $R_I$\\
1 & $5.778\times10^{-7}$ & $1.41\times10^{-17}$ & $-6.99\times10^{-7}$ \\
2 & $-3.057\times10^{-5}$ & $2.92\times10^{-17}$ & $-1.32\times10^{-5}$ \\
3 & $-4.800\times10^{-5}$ & $7.86\times10^{-18}$ & $-1.19\times10^{-5}$ \\
\end{tabular}
\end{ruledtabular}
\end{table}

On a vacuum Kasner trajectory $p_\psi$ is constant. During the nonlinear event the instantaneous velocities need not satisfy the free constraint, so the $A_0$ contribution evaluated at that state can be nonzero. Keeping it avoids attributing a geometric normalization effect to spatial transport. The decomposition describes a measured solution. A causal comparison in which one source is removed would require solving another consistent system.

\begin{table*}[t]
\caption{Central first-exchange integrals from the first magnetic maximum through two pulse widths. The width is the connected full width at half maximum in volume time, giving a different window for each input. Table~\ref{sm:tab:closure} gives the vacuum and remaining matter contributions and temporal quadrature residuals. The latter are at most $1.4\times10^{-5}$; evolution error is estimated separately.}
\label{sm:tab:pulse}
\begin{ruledtabular}
\begin{tabular}{ccccccc}
Input & $s_{\rm peak}$ & $T$ interval & $I_E$ & $I_B$ & $I_\nabla$ & $\Delta p_\psi$\\
1 & 285.1 & 259.24--273.48 & +0.02515 & -0.04976 & +0.01385 & -0.01075 \\
2 & 276.6 & 252.24--268.00 & +0.04828 & -0.03887 & +0.00811 & +0.01750 \\
3 & 269.9 & 246.72--262.72 & +0.05003 & -0.03680 & +0.00669 & +0.01988 \\
\end{tabular}
\end{ruledtabular}
\end{table*}

\section{The Isolated Magnetic-Wall Limit}
\label{sm:app:wall}

Take a homogeneous diagonal spatial metric with scale factors $e^{-\beta_a}$ and lapse equal to the spatial volume, up to the fixed normalization of Hamiltonian time $\zeta$. Set $\varphi=\sqrt2\psi$. The untruncated momentum and off-diagonal Einstein equations require $\mathbf E\mathbin{\times}\mathbf B=0$ and $E_aE_b+B_aB_b=0$ for $a\ne b$, with fields in an orthonormal frame. A single magnetic axis, chosen as axis 2, with all electric components zero satisfies these conditions. Absorbing constant flux factors into $B^2$, its exact Hamiltonian is
\begin{align}
 H={}&\frac12\left[\sum_a\pi_a^2-\frac12\left(\sum_a\pi_a\right)^2+\pi_\varphi^2\right]\notag\\
 &+\Omega_B=0,\notag\\
 \Omega_B&=B^2e^{-2\beta_2+\varphi^2/2}.
 \label{sm:eq:wallhamiltonian}
\end{align}
This follows from the inverse DeWitt metric on logarithmic scale factors and the conserved Maxwell magnetic flux. The independent scalar kinetic term fixes the normalization of the Kasner constraint. Hamilton's equations are
\begin{align}
 \beta_a'&=\pi_a-\tfrac12\sum_b\pi_b,&\varphi'&=\pi_\varphi,\notag\\
 \pi_a'&=2\delta_{a2}\Omega_B,&
 \pi_\varphi'&=-\varphi\Omega_B,
 \label{sm:eq:hamiltoneq}
\end{align}
where primes denote $\dd/\dd\zeta$. In free flight, $\sum_a\beta_a'=1$ fixes the incoming normalization and $\pi_a=p_a-1$, $\pi_\varphi=p_\psi$.

If the wall potential has initial value $W$, initial data satisfying the constraints in this single-axis sector and continuously connected to a free Kasner state follow from the common spatial momentum shift
\begin{equation}
 \pi_a=p_a-1+\eta,
 \qquad\eta=\frac23(1-\sqrt{1+3W}),\qquad
 \pi_\varphi=p_\psi.
 \label{sm:eq:wallinitial}
\end{equation}
Indeed the kinetic constraint changes by $\eta-3\eta^2/4=-W$. The volume rate is positive on this branch. The instantaneous normalized state is the four-vector $(\beta_1',\beta_2',\beta_3',\varphi')/\sum_a\beta_a'$. In particular Hamiltonian time and logarithmic volume time coincide on the normalized free trajectory, and their relation during the wall is obtained by integrating the volume rate.

For one magnetic wall and positive incoming scalar, let $\varphi_*$ be the scalar at the actual magnetic maximum and set $q_B=\varphi_*/2$. Define $\epsilon=\sqrt\delta$ and the inner variables
\begin{align}
 z&=q_B\epsilon(\zeta-\zeta_{\rm peak}),\qquad
 \chi=q_B(\varphi-\varphi_*),\notag\\
 \widehat\pi_\varphi&=\pi_\varphi/\epsilon,\qquad
 \widehat\Omega_B=\Omega_B/\epsilon^2.
 \label{sm:eq:innercoords}
\end{align}
For $q_B\gg1$ and $\epsilon q_B\ll1$, the leading equations across the isolated magnetic layer are
\begin{equation}
 \partial_z\widehat\pi_\varphi=-2\widehat\Omega_B,\quad
 \partial_z\chi=\widehat\pi_\varphi,\quad
 \partial_z\ln\widehat\Omega_B=2\widehat\pi_\varphi.
 \label{sm:eq:liouville}
\end{equation}
For two free epochs satisfying the quantitative matching conditions below, the conditional tangent reflection is
\begin{align}
 \widehat P_+&=\mathcal R_\infty\widehat P_-+O(q_B^{-1}),\notag\\
 \mathcal R_\infty(\widehat p_a;\widehat p_\psi)
 &=(\widehat p_a;-\widehat p_\psi),\qquad
 \widehat P=(P-\PT)/\sqrt\delta,
 \label{sm:eq:reflection}
\end{align}
where $\PT=(0,0,1;0)$. The width in $\epsilon\zeta$ and the spatial impulse divided by $\epsilon$ are $O(q_B^{-1})$. The homogeneous layer thus provides a reflection law for an isolated interaction.

If $\Omega_{B,K}=|\mathcal A_B|^2/2$ on the incoming surface, Eq.~\eqref{sm:eq:wallhamiltonian} gives the exact identity
\begin{align}
 q_B^2={}&L_B+\tfrac12\ln(2\Omega_{B,*})
 +\Delta\beta_{\rm wall}+\tfrac14\varphi_K^2,\notag\\
 \Delta\beta_{\rm wall}&=\beta_{{\rm wall},*}-\beta_{{\rm wall},K}.
 \label{sm:eq:peakslope}
\end{align}
The free-flight threshold in Eq.~\eqref{sm:eq:tstar} sets a wall proxy to unity. The magnetic maximum in the critical layer instead has $\Omega_{B,*}=O(\delta)$. Its logarithm can therefore compete with $L_B$ for an algebraically vanishing seed. The actual peak-clock coefficient and the relation between $q_B$ and $L_B$ require matching this energy and the spatial displacement; they are not fixed by the unit-energy threshold.

Expanding about the actual peak gives
\begin{equation}
 \frac{\varphi^2-\varphi_*^2}{2}=2\chi+\frac{\chi^2}{2q_B^2},
 \qquad q_B=\varphi_*/2.
 \label{sm:eq:wallexponent}
\end{equation}
The scalar Hamilton equation approaches Eq.~\eqref{sm:eq:liouville}. The exact logarithmic derivative of the magnetic term additionally contains $-2\beta_2'$. Over a fixed inner interval this contributes $O(q_B^{-2})+O(\epsilon/q_B)$ in the rescaled equation: the wall-induced spatial velocity is $O(\epsilon/q_B)$, and the incoming velocity along the magnetic axis is $O(\epsilon^2)$. With the wall centered at $z=0$, the leading solution is
\begin{align}
 \widehat\pi_\varphi&=-p\tanh(pz),\notag\\
 \widehat\Omega_B&=\frac{p^2}{2}\sech^2(pz),\qquad
 \chi=-\ln\cosh(pz),
 \label{sm:eq:wallsolution}
\end{align}
which conserves $\widehat\pi_\varphi^2+2\widehat\Omega_B=p^2$. The coefficient $p$ is the incoming scalar velocity in the rescaled free normalization. An exact impulse estimate connects this core to the free epochs.

\subsection{Finite-interval impulse and matching}

Write $u=\pi_\varphi$, $V=\sum_a\beta_a'$ and $a=\beta_{2,-}'$. On an interval $[\zeta_-,\zeta_+]$ containing the magnetic maximum, let
\begin{equation}
 I(\zeta)=\int_{\zeta_-}^{\zeta}\Omega_B\,\dd\zeta.
 \label{sm:eq:wallimpulse}
\end{equation}
The exact equations imply $\beta_2'=a+I$, $\beta_i'=\beta_{i,-}'-I$ for $i\ne2$, and $V=V_--I$. In particular,
\begin{align}
 (u^2+2\Omega_B)'&=-4\beta_2'\Omega_B,\notag\\
 u^2+2\Omega_B&=u_-^2+2\Omega_{B,-}-4aI-2I^2.
 \label{sm:eq:wallexactenergy}
\end{align}
This identity includes the geometric work throughout the event.

To quantify isolation, take constants independent of $q_B$ and $\epsilon$ such that $c\epsilon\leq u_-\leq C\epsilon$, $0\leq a\leq C\epsilon^2$, and $0<v_0\leq V_-\leq v_1$. Require $\varphi\geq q_B$ on the matching interval and small endpoint energies,
\begin{equation}
 \eta_B=\frac{\Omega_{B,-}+\Omega_{B,+}}{\epsilon^2}\ll c^2.
 \label{sm:eq:wallendpoints}
\end{equation}
The outgoing endpoint lies on the decreasing magnetic branch. These conditions concern an isolated finite event and its adjacent free portions.

Equation~\eqref{sm:eq:wallexactenergy} bounds $|u|$ by a constant times $\epsilon$. Integrating $u'=-\varphi\Omega_B$ then gives
\begin{equation}
 0\leq I_+\leq\frac{u_-+\sup|u|}{q_B}
 =O(\epsilon/q_B).
 \label{sm:eq:impulsebound}
\end{equation}
Hence $V$ stays bounded away from zero. The energy identity at the outgoing endpoint yields
\begin{equation}
 u_+^2-u_-^2
 =O(\epsilon^2\eta_B+\epsilon^3/q_B+\epsilon^2/q_B^2).
 \label{sm:eq:wallenergybound}
\end{equation}
For sufficiently small $\eta_B$ and large $q_B$, this bounds $|u_+|$ below by $c\epsilon/2$. If $u_+\geq0$, the decreasing-wall condition would instead require $u_+<2(a+I_+)/\varphi_+=O(\epsilon^2/q_B+\epsilon/q_B^2)$, a contradiction. Thus the scalar reverses, and
\begin{equation}
 |u_++u_-|=O\!\left[\epsilon(\eta_B+\epsilon/q_B+q_B^{-2})\right].
 \label{sm:eq:scalarreversalbound}
\end{equation}
With $s_a=(-1,1,-1)$, spatial normalization is controlled exactly by
\begin{equation}
 p_{a,+}-p_{a,-}
 =\frac{I_+(s_a+p_{a,-})}{V_--I_+}.
 \label{sm:eq:wallvolumenormalization}
\end{equation}
The scalar normalization $u/V$ contributes a further $O(\epsilon/q_B)$ to the tangent error. If $\nu$ bounds the uncertainty of the two extracted spectra after division by $\epsilon$, Eqs.~\eqref{sm:eq:impulsebound}--\eqref{sm:eq:wallvolumenormalization} give
\begin{equation}
 \mathcal D_B\equiv\frac{d_K(P_+,\mathcal R_\infty P_-)}{\epsilon}
 \leq C(q_B^{-1}+\eta_B)+\nu.
 \label{sm:eq:uniformwallbound}
\end{equation}
Here the reflection acts by reversing the scalar component; its action on $\widehat P$ is Eq.~\eqref{sm:eq:reflection}. The constant is uniform under the stated interval bounds.

For matching to Liouville tails, choose both endpoint levels $\Omega_{B,\pm}/\epsilon^2=c_\pm q_B^{-2}$, with positive constants $c_\pm$ independent of $q_B$ and $\epsilon$. The leading solution reaches these levels at $|z|=O(\ln q_B)$, so the matching interval grows with $q_B$. On either monotone tail below a fixed small fraction of $\epsilon^2$, Eq.~\eqref{sm:eq:wallexactenergy} keeps $|u|\geq c_1\epsilon$. The exact equation $(\ln\Omega_B)'=\varphi u-2\beta_2'$ therefore gives $|(\ln\Omega_B)'|\geq c_2q_B\epsilon$ and
\begin{equation}
 \int_{\rm tail}\Omega_B\,\dd\zeta
 \leq\frac{\Omega_{B,\rm edge}}{c_2q_B\epsilon}.
 \label{sm:eq:walltail}
\end{equation}
The corresponding scalar excursion is $O(\ln q_B/q_B)$. The exponent correction in Eq.~\eqref{sm:eq:wallexponent} is then $O(\ln^2q_B/q_B^2)$, while the accumulated scalar-force correction obeys
\begin{equation}
 \left|\int(\varphi-2q_B)\Omega_B\,\dd\zeta\right|
 \leq O(\ln q_B/q_B)I_+
 =O(\epsilon\ln q_B/q_B^2).
 \label{sm:eq:wallweightedremainder}
\end{equation}
The geometric work $4aI_++2I_+^2$ and the volume correction have already been retained exactly. These finite-tail estimates close the passage from the core to the normalized free spectra. For $\eta_B+\nu=O(q_B^{-1})$, Eq.~\eqref{sm:eq:uniformwallbound} gives Eq.~\eqref{sm:eq:reflection}.

Additional compatible sources can be retained explicitly when their effect takes the form $\beta_a''=s_a\Omega_B+f_a$, $u'=-\varphi\Omega_B+g_\varphi$, with the same conserved magnetic flux. Define
\begin{align}
 \mathfrak F&=\epsilon^{-1}\sum_a\int|f_a|\dd\zeta,\notag\\
 \mathfrak G&=\epsilon^{-1}\int|g_\varphi|\dd\zeta,\qquad
 F_2(\zeta)=\int_{\zeta_-}^{\zeta}f_2\dd\zeta,\notag\\
 \mathfrak W&=\epsilon^{-2}\int|2ug_\varphi-4F_2\Omega_B|\dd\zeta.
 \label{sm:eq:wallotherforces}
\end{align}
The bounded-energy estimate acquires the explicit remainder
\begin{equation}
 \mathcal D_B\leq C(q_B^{-1}+\eta_B+\mathfrak F+\mathfrak G+\mathfrak W)+\nu.
 \label{sm:eq:wallremainderbound}
\end{equation}
The same interval bounds and small accumulated remainders are required. The abbreviated $O(q_B^{-1})$ statement applies only when their sum is $O(q_B^{-1})$. The exact single-axis controls below have $f_a=g_\varphi=0$.

\subsection{Self-consistent finite controls}

Eight local controls use the measured Kasner parameters at $\delta=10^{-6},10^{-8}$ with $\varphi_K=10,20,40,80$. Initially all $\beta_a=0$, $\Omega_{B,K}=10^{-7}\delta$, and all electric fields vanish; Eq.~\eqref{sm:eq:wallinitial} fixes the momenta. These choices specify local homogeneous data. The actual $q_B$ is obtained from a zero of $(\ln\Omega_B)'$ on the descending branch, using continuous integration output.

The equations are integrated with DOP853 at relative tolerances $10^{-9},10^{-11},10^{-13}$ and absolute tolerances one hundredth as large. Radau independently evolves the exact variables $x=\epsilon\zeta$, $(\pi_a-\pi_{a,T})/\epsilon$, $u/\epsilon$, with $\pi_{a,T}=(-1,-1,0)$, at relative tolerance $10^{-11}$. Both formulations retain the entire single-wall Hamiltonian and use $\ln\Omega_B$, avoiding exponentially large coupling factors.

Each spectrum is averaged on a window at least one magnetic full width at half maximum (FWHM) long in logarithmic volume time. Endpoint potential levels $\Omega_B/\Omega_{B,K}=1,0.1,0.01$, followed by extension by another FWHM, test extraction dependence on both sides. Across the eight controls, full-spectrum numerical differences divided by $\epsilon$ are below $4.29\times10^{-10}$, window differences below $1.80\times10^{-9}$, and plateau drift and finite-window matter effects below $1.36\times10^{-8}$. Each is below one tenth of the resolved $\mathcal D_B$.

\clearpage
\begin{table}[t]
\caption{Pure single-axis controls illustrating the analytic reflection bound. $\mathcal D_B$ is the tangent discrepancy in Eq.~\eqref{sm:eq:uniformwallbound}; $u_B$ is the largest of the numerical, window, drift and finite-window matter estimates in the same units.}
\label{sm:tab:purewall}
\begin{ruledtabular}
\begin{tabular}{cccccc}
$\delta$ & $\varphi_K$ & $q_B$ & $\mathcal D_B$ & $q_B\mathcal D_B$ & $10^8u_B$\\
$10^{-6}$ & 10 & 6.028 & 8.635 & 52.053 & 1.35 \\
$10^{-6}$ & 20 & 10.551 & 4.936 & 52.079 & 1.21 \\
$10^{-6}$ & 40 & 20.281 & 2.567 & 52.069 & 1.15 \\
$10^{-6}$ & 80 & 40.141 & 1.297 & 52.054 & 1.12 \\
$10^{-8}$ & 10 & 6.027 & 8.620 & 51.949 & 1.35 \\
$10^{-8}$ & 20 & 10.551 & 4.932 & 52.035 & 1.21 \\
$10^{-8}$ & 40 & 20.281 & 2.567 & 52.064 & 1.14 \\
$10^{-8}$ & 80 & 40.141 & 1.297 & 52.071 & 1.11 \\
\end{tabular}
\end{ruledtabular}
\end{table}

The four-dimensional equations provide a direct check of this field sector. Multiplication by $N^2$ gives $R_{\hat0\hat0}N^2=V'+V^2-\sum_a\beta_a'^2$ and $R_{\hat a\hat a}N^2=-\beta_a''$. Their matter sources are $u^2+\Omega_B$ and $-s_a\Omega_B$, respectively. The scalar equation is $u'=-\varphi\Omega_B$; the homogeneous magnetic flux satisfies Maxwell's equations and the Bianchi identity. Direct differentiation of the metric verifies these formulas. Independently differencing the stored geometric velocities and scalar momentum gives approximately fourth-order residual convergence. These checks supplement the integration comparison and do not replace its output error estimate.

A separate 495-trajectory Hamiltonian ensemble spans eleven backgrounds, three amplitude scalings and fifteen frequency/polarization choices. Its main integration includes a perpendicular electric term $D_3^2e^{-2\beta_3-\varphi^2/2}$, initialized at fraction $10^{-8}$; its comparison integration evolves a single magnetic potential without that electric term. The recorded Hamiltonian residual is at most $1.14\times10^{-13}$, the relative map difference $9.18\times10^{-8}$, and the peak-clock difference $1.04\times10^{-7}$. A perpendicular electric field sources a momentum equation absent from the diagonal homogeneous ansatz. These are restricted Hamiltonian comparisons; the self-consistent four-dimensional reference and its conditional error estimate are supplied by the single-axis analysis above.

Spatial axes are compared with
\begin{equation}
 d_K(P,Q)=\min_{\pi\in S_3}
 \left[\sum_{a=1}^3(p_a-q_{\pi(a)})^2+(p_\psi-q_\psi)^2\right]^{1/2}.
 \label{sm:eq:distance}
\end{equation}
The physical magnetic axis is retained in the Hamiltonian before the spectra are ordered. A propagating solution with two plateaus satisfying the criteria below can be compared with the isolated wall using its own incoming geometric state, scalar and magnetic field. The finite observation windows for the packets do not supply that pair of plateaus.

\section{Regular Linear Horizon Data and the Local Model}
\label{sm:app:horizon}

The static black-hole backgrounds motivate the local parameters, and their coupled odd-parity channels supply the carrier's complex field and momentum. Gauge-invariant master equations and reconstruction in general EMS theory provide an independent reference for this linear calculation~\cite{sm:Jansen2019}. The even-parity sector contains three coupled physical master fields, which become relevant to the response generated by an odd wave at second order.

\subsection{Carrier basis and matching surface}

The carrier uses the two $\ell=2$ odd modes in Regge--Wheeler gauge. In a spherical background $\dd s^2=-N_{\rm BH}e^{-2\chi}\dd t_{\rm BH}^2+\dd r_{\rm BH}^2/N_{\rm BH}+r_{\rm BH}^2\dd\Omega^2$, put $z=1/r_{\rm BH}$ and $h=N_{\rm BH}e^{-\chi}/z$. The canonical basis $\mathbf V=(V_g,V_e)^{\mathsf t}$ is fixed by
\begin{equation}
 \mathfrak q=\frac{e^{-\chi}z}{\sqrt2}V_g,
 \qquad A_v=\frac{V_e}{\sqrt{8[\ell(\ell+1)-2]Z}},
 \label{sm:eq:carrierbasis}
\end{equation}
where $\mathfrak q$ is the odd gravitational auxiliary field and $A_v$ the odd Maxwell harmonic amplitude. These are the canonical variables of Gannouji and Rodr\'iguez Baez~\cite{sm:Gannouji2022}, specialized to $f_1=2$, $f_2=4X+4ZF$ and their charge $4Q_{\rm BH}$. The radial system is
\begin{equation}
 \partial_{r_*}\mathbf V=\boldsymbol\Pi,\qquad
 \partial_{r_*}\boldsymbol\Pi=(\mathbf V_{\rm odd}-\omega^2\mathbf1)\mathbf V,
 \label{sm:eq:carriertransfer}
\end{equation}
with $\dd r_*/\dd r_{\rm BH}=e^{\chi}/N_{\rm BH}$ and the symmetric odd potential $\mathbf V_{\rm odd}$. Integration begins at $z=1+10^{-10}$, where $\mathbf V=\mathbf1/\sqrt\omega$ and $\boldsymbol\Pi=-i\sqrt\omega\,\mathbf1$ fix the positive-frequency ingoing normalization and phase. Thus the symplectic flux is $-\mathbf1$. We set $r_H=1$, $\chi_\infty=0$, and $\omega=0.2/M$.

The matching surface $x_K=\ln(1/r_K)$ is the first surface on which the Kasner residual is below $10^{-6}$ and remains so over a unit interval of $x=\ln z$. The local conversion is $\Lambda=a_\parallel|\tr K|$, so $k=\omega/\Lambda$ and $w=\Lambda M\tau_0/\sqrt\delta$. With the initial orbit scales set to unity, the reduced charge is $q=4Q_{\rm BH}e^{2x_K}/|\tr K|$. The matching parameters, canonical transfer matrices and momenta appear in Table~\ref{sm:tab:canonicaltransfer}.

\begin{table*}[t]
\caption{Carrier matching parameters and complex field and momentum matrices. $x_K=\ln(1/r_K)$, $M$ is the ADM mass in units $r_H=1$, and $\Lambda=a_\parallel|\tr K|$. Matrix rows are ordered as $(V_g,V_e)$; columns label the two ingoing channels.}
\label{sm:tab:canonicaltransfer}
\begin{ruledtabular}
\begin{tabular}{ccccc}
$\delta$ & $M$ & $x_K$ & $\beta_K$ & $\Lambda$\\
$10^{-6}$ & 0.901148779 & 0.225070276 & 76.84086557 & 0.155115119 \\
$10^{-5}$ & 0.901214604 & 0.259907636 & 24.36647729 & 0.166808659 \\
\end{tabular}
\end{ruledtabular}
\begin{ruledtabular}
\begin{tabular}{ccc}
$\delta$ & $\mathbf V_K$ & $\boldsymbol\Pi_K$\\
$10^{-6}$ & $\begin{pmatrix}-1.351901+0.221835i & 1.289882-0.982632i\\1.293612-0.978047i & 0.599001-1.258915i\end{pmatrix}$ & $\begin{pmatrix}0.058324+0.300143i & -0.218499-0.284207i\\-0.217427-0.284972i & -0.270867-0.129721i\end{pmatrix}$ \\
$10^{-5}$ & $\begin{pmatrix}1.132605+0.811154i & -1.606383-0.228911i\\-1.598256-0.297249i & -1.308844+0.410959i\end{pmatrix}$ & $\begin{pmatrix}0.173571-0.244893i & -0.052008+0.354794i\\-0.067098+0.351568i & 0.082650+0.293537i\end{pmatrix}$ \\
\end{tabular}
\end{ruledtabular}
\end{table*}

Let $\lambda_h=\ell(\ell+1)$, $n_h=\sqrt{8(\lambda_h-2)}$, and $N_{\rm BH}<0$ on the interior surface. The compatible orthonormal magnetic and electric rows, normalized by $\sqrt{K_{ij}K^{ij}}$, are
\begin{align}
 \mathbf b_K&=\frac{-i\omega\sqrt{\lambda_h}e^\chi z}
 {n_h\sqrt{-N_{\rm BH}}\sqrt{K_{ij}K^{ij}}}(\mathbf V_K)_{e,:},\notag\\
 \mathbf e_K&=\frac{\sqrt{\lambda_h}\sqrt{-N_{\rm BH}}}
 {h n_h\sqrt{K_{ij}K^{ij}}}
 [ (\boldsymbol\Pi_K)_{e,:}+S_2(\mathbf V_K)_{e,:}],\notag\\
 S_2&=\tfrac12h z^2\beta\,\partial_\psi\ln Z.
 \label{sm:eq:carrierrows}
\end{align}
The harmonic convention is root-mean-square normalization, with the local transverse axis aligned with the odd vector harmonic. At $\delta=10^{-6}$ choose $\mathbf c_H=\mathbf b_K^*/\|\mathbf b_K\|_2$, which maximizes $|\mathbf b_K\mathbf c_H|$ at $\|\mathbf c_H\|_2=1$. Numerically,
\begin{equation}
 \mathbf c_H=\begin{pmatrix}-0.45732712+0.60488269i\\-0.58865900+0.28008824i\end{pmatrix}.
 \label{sm:eq:fixedpolarization}
\end{equation}
This same polarization is used on both backgrounds. The $B,E$ in Table~\ref{sm:tab:initialparams} are $(\mathbf b_K\mathbf c_H,\mathbf e_K\mathbf c_H)\sqrt{K_{ij}K^{ij}}/|\tr K|$. The conjugate data and the envelope derivative therefore fix the local electric field independently of its magnetic amplitude.

The background sequence gives the estimate for $c_0$. Its three smallest recorded distances yield $\beta\sqrt\delta=0.0768169335$, $0.0768169997$, and $0.0768171519$ at $\delta\simeq10^{-9},10^{-8.5},10^{-8}$, respectively. The value $0.076817$ summarizes these points; the propagation comparison uses the measured parameters at each finite background.

\subsection{Regular low-frequency fields}

For a regular normalization continuous through zero frequency, write the two-component master solution as
\begin{equation}
 \mathbf v=e^{-i\omega r_*}\mathbf H(x),\qquad
 f(x)=\frac{\dd x}{\dd r_*}=f_1x+f_2x^2+\cdots,
 \label{sm:eq:regularhorizon}
\end{equation}
where $x=0$ is the horizon and the matrix potential is $\mathbf V(x)=\mathbf V_1x+O(x^2)$. Factoring out the ingoing phase gives
\begin{equation}
 f^2\mathbf H_{xx}+(ff_x-2i\omega f)\mathbf H_x-\mathbf V\mathbf H=0.
 \label{sm:eq:regularode}
\end{equation}
Choose $\mathbf H(0)=\mathbf1$ and fix the tortoise constant by $r_*-f_1^{-1}\ln x\to0$. The first regular coefficient and the conjugate data are
\begin{align}
 \mathbf H_x(0)&=\frac{\mathbf V_1}{f_1(f_1-2i\omega)},\notag\\
 \boldsymbol\Pi&=e^{-i\omega r_*}(f\mathbf H_x-i\omega\mathbf H).
 \label{sm:eq:regularmomentum}
\end{align}
This horizon-field normalization has a continuous $\omega=0$ limit for both channels. Static and nonstatic data use the same basis. The positive-frequency carrier above instead has unit symplectic flux.

The low-frequency checks find a finite master-field limit. The reconstructed compatible magnetic amplitude is nevertheless proportional to $\omega$ near zero frequency on the reference background. Consequently a low-frequency local seed cannot be assigned a horizon amplitude by identifying it with a finite static master field. Real packet matching requires the full frequency-dependent field and momentum, including the actual source of the added component in Eq.~\eqref{sm:eq:packetexplicit}.

There is also a geometric restriction. In a strict two-Killing Maxwell field, the Bianchi identity forces $\partial_TF_{AB}=\partial_yF_{AB}=0$. A spherical odd harmonic generally has a nonconstant transverse flux and a metric perturbation in addition to the compatible local components used here. The local model with $F_{AB}=0$ therefore does not reproduce all components of that harmonic. A full black-hole construction would retain the induced even-parity fields, solve their constraints with no independent second-order incoming radiation and zero asymptotic scalar source, and match their conjugate data in an overlap region. Angular residuals and their forced response would then test the spatial approximation.

The initial black-hole mass must be held fixed in such a construction. Charge conservation and positive absorbed energy would change the critical distance according to
\begin{equation}
 \delta_f=\frac{1+\delta_i}{1+\Delta M/M_i}-1.
 \label{sm:eq:massdrift}
\end{equation}
If a separately normalized horizon family has $\Delta M/M_i=O(\delta^{1/2+2\kappa})$, its relative mass-induced critical drift is $O(\delta^{2\kappa-1/2})$. This is conditional power counting, not a measured absorption law for the local equal-energy packets. At $\kappa=1/2$ it suggests a parametrically small drift, while $\kappa=1/4$ would require direct competition with the critical distance.

\section{Numerical Methods and Validity}
\label{sm:app:numerics}

\subsection{Evolution and domain}

The main discretization uses centered fourth-order spatial derivatives and fourth-order Runge--Kutta time stepping on the constrained nonlinear system. The compact initial envelope is exactly zero outside $|y|<w$. The computation domain includes the causal past of every observation curve throughout its time window; periodic wraparound is excluded. Enlarging the domain is tested separately. Constraint-compatible initial $\sigma$ is integrated from the momentum source, and its time derivative is determined by the remaining constraint.

At $\delta=10^{-5}$ the spatial spacings are $w/128,w/256,w/512$; independent time-step comparisons at spacing $w/256$ use $\Delta T/h=0.4,0.2,0.1$. Final full-spectrum spatial convergence orders are about $3.98$ for both inputs, with middle-to-fine differences below $1.93\times10^{-9}$. Time-step convergence orders are $3.94$--$3.96$, with the finest difference below $3.82\times10^{-11}$. The largest fine-grid constraint residual, normalized by squared expansion, is $6.14\times10^{-11}$. Doubling the domain leaves the observed spectra identical in double precision.

The reference-background calculations use spacings $0.4,0.2,0.1$, with independent time stepping and domain tests on the decisive larger response. At the early comparison time, changing the carrier phase is checked on all three grids. The three reference trajectories continue through $T=380$. The maximum middle-to-fine full-spectrum difference over that interval is below $3.34\times10^{-6}$ and is below $1.38\times10^{-7}$ through $T=300$. These numerical estimates apply to the local EMS evolution.

The initial continuous energy is evaluated by Gaussian quadrature of the analytic compact profiles. Finite-difference energies converge at fourth order; at the finest $\delta=10^{-5}$ grid their relative offset from the continuum target is about $7.86\times10^{-6}$. The offset is shared by the two inputs to a precision far better than the measured response difference. Tiny low-frequency energies are computed through Eq.~\eqref{sm:eq:energycoefs}, so that their meaning does not rely on subtraction at machine precision.

\subsection{Prediction error versus evolution error}

All three comparisons use the same preselected times and normal curves for the coherent prediction, the prediction without cross terms and nonlinear evolution. The numerical estimate includes spatial and time refinement and temporal interpolation tested by subsampling the histories. The expansion discrepancy is reported separately. Table~\ref{sm:tab:errors} summarizes these quantities for the larger admixture at $\delta=10^{-5}$. The conservative numerical estimate is less than one tenth of the difference resolved in every comparison.

\begin{table}[!htbp]
\caption{Prediction discrepancies for $\lambda_3$, $\delta=10^{-5}$. The first two error columns are percentages normalized by the maximum coherent profile. $\epsilon_{\rm num}$ is the combined numerical estimate in absolute $p_\psi$. It does not include the amplitude-expansion discrepancy.}
\label{sm:tab:errors}
\begin{ruledtabular}
\begin{tabular}{cccc}
$T$ & Coherent & No cross & $\epsilon_{\rm num}$\\
53.395 & 0.38 & 46.90 & $1.44\times10^{-8}$ \\
54.061 & 1.12 & 47.52 & $1.64\times10^{-8}$ \\
54.749 & 3.48 & 49.75 & $1.34\times10^{-7}$ \\
\end{tabular}
\end{ruledtabular}
\end{table}

The controls comprise evolution with no packet, an isolated homogeneous wall, and constant coupling $Z=1$. Constant coupling removes the direct scalar source from the electromagnetic field; its maximum spectrum change in the long reference comparison is below $2.95\times10^{-10}$. This change of physical equations isolates the role of the exponential coupling in the observed response.

\subsection{Independent discretization and field equations}

For the comparison between backgrounds, sixth-order spatial differences with DOP853 time integration evolve from the initial surface. The maximum difference from the fine fourth-order calculation is $1.29\times10^{-10}$ in the full geometric spectrum over the compared histories. Both evolutions begin from the same constrained initial fields and share no evolved intermediate state. For the long reference calculation, the independent evolution covers the interaction interval from a shared state near $T=220$ through $T=300$, with a finest spectrum difference below $7.5\times10^{-9}$. Its coverage is confined to this interval; earlier propagation and the continuation to $T=380$ use the resolution tests above.

A separate Maxwell calculation in continuous frequency retains the complex amplitudes and momenta. Simultaneously halving the frequency spacing, increasing the cutoff and refining initial quadrature changes the combined fields by less than $6.28\times10^{-10}$ relatively in the comparison at $\delta=10^{-5}$. Its discrepancy with the spatially discretized electromagnetic fields is below $6.3\times10^{-8}$; the estimated truncated tail is of order $10^{-12}$. High-precision evaluation of the compact-envelope transform also checks the small spectral tail.

The four-dimensional fields, including the orbit connection, are reconstructed and inserted into Eq.~\eqref{sm:eq:fourdim} using local derivative fits. For the fine samples at $\delta=10^{-5}$, the relative Einstein and Maxwell residuals are below $9.21\times10^{-10}$ and $1.43\times10^{-10}$; the scalar residual divided by squared expansion is below $8.86\times10^{-13}$. The absolute reduced-energy residual is below $9.35\times10^{-12}$, and the scalar and transverse electromagnetic balances are checked separately. These small residuals check consistency with the field equations. Their variation under spatial and derivative-fit refinement is not uniformly monotone, so they do not establish a converged four-dimensional reconstruction error or control of omitted angular dynamics. Output errors are estimated from the converged observables.

\subsection{Kasner plateaus}

The spectrum is extracted from the reconstructed spatial metric and extrinsic curvature, in an orbit-adapted orthonormal frame. This avoids losing a small base metric to cancellation against the connection terms. The scalar carries the factor $\sqrt2$ in Eq.~\eqref{sm:eq:spectrum}; only spatial eigenvalues may be permuted. For the comparison at $\delta=10^{-5}$, the difference between reconstructed and polarized diagonal eigenvalue formulas is below $1.2\times10^{-16}$.

A Kasner plateau must last at least one resolved magnetic pulse width in logarithmic volume time. Over that interval the full-spectrum drift and the integrated absolute matter and spatial contributions must each be at most $10^{-4}$; normalized matter and spatial fractions must also be at most $10^{-4}$. Numerical error must remain below $10^{-5}$, and these conditions must persist when the observation window is extended. Through $T=340$ and $380$, none of the fifteen monitored curves has an outgoing plateau satisfying these conditions. A later plateau remains possible. The finite exchange therefore has no assigned scattering output or relative single-wall scattering error.
\section{Additional Response and Exchange Comparisons}

All reference inputs use $\delta=10^{-6}$, $\kappa=1/2$, $\omega M=0.2$, $\tau_0=1$ and the same reference polarization. In addition to $\lambda_1=8.103112480050155\times10^{-7}$ and $\lambda_3=1.78726246450643\times10^{-5}$, the intermediate input is $\lambda_2=4.910111501785908\times10^{-6}$. They were selected to give predicted relative responses $0.01$, $0.1$ and $1$ at $T=242.44843145$, which is $1.4$ times the fixed reference magnetic clock. The envelope in Eq.~\eqref{sm:eq:packetexplicit} is $f(u)=\exp[1-(1-u^2)^{-1}]$ for $|u|<1$, and zero otherwise. The initial transverse energies are $7.45941\times10^{-10}$ at $\delta=10^{-6}$ and $2.35513\times10^{-8}$ at $\delta=10^{-5}$.

\begin{table}[t]
\caption{Reference comparison at $T=242.4484$ using the same 175 curves in $|y|\leq34.8$ for all columns. Responses and profile discrepancies are percentages. The last row lies beyond quantitative second order, where even the prediction without cross terms has a smaller discrepancy than the coherent approximation.}
\label{sm:tab:reference}
\begin{ruledtabular}
\begin{tabular}{ccccc}
Input & $100A_\psi^{(2)}$ & $100A_\psi$ & Coherent & No cross\\
1 & 1.000 & 0.995 & 0.52 & 48.38 \\
2 & 10.000 & 9.505 & 4.95 & 24.68 \\
3 & 100.000 & 65.772 & 34.23 & 23.44 \\
\end{tabular}
\end{ruledtabular}
\end{table}

On the changed background, $\lambda_1$ and $\lambda_3$ retain their values and phase $\vartheta=0$. Every comparison uses the same 65 curves normal to the time slices in $|y|\leq w/4$. The first two times mark predicted $1\%$ and $3\%$ response for $\lambda_1$; the third is the last common sampled prediction time below $10\%$ in the prescribed observation window. The nonlinear ordering is reversed at all three times.

\begin{table}[t]
\caption{Signed parts of $100\Delta p_\psi^{(2)}/|\pzero|$ at $y=-w/4$ for $\lambda_3$, $\delta=10^{-5}$. Each term includes its initial constraints and induced geometry. The cross term partly cancels the carrier response.}
\label{sm:tab:sources}
\begin{ruledtabular}
\begin{tabular}{crrrr}
$T$ & Carrier & Cross & Low freq. & Total\\
53.395 & -1.0149 & +0.3309 & -0.0271 & -0.7111 \\
54.061 & -3.0447 & +0.9906 & -0.0808 & -2.1350 \\
54.749 & -9.6983 & +3.1492 & -0.2564 & -6.8055 \\
\end{tabular}
\end{ruledtabular}
\end{table}

The three reference inputs produce first magnetic maxima at central volume times $s\simeq285.1,276.6,269.9$. The Maxwell damping coefficient $2\psi\psi_T-P_T$ becomes negative during the central interaction, allowing the electric field to grow after the scalar turns. Through $T=380$, none of the fifteen monitored curves, five for each input, has an outgoing plateau satisfying the stated criteria. A later plateau remains possible.
\begingroup
\renewcommand{\bibsection}{\section*{References}}
\let\SupplementLabel\label
\let\SupplementRef\ref
\renewcommand{\label}[1]{\SupplementLabel{sm:#1}}
\renewcommand{\ref}[1]{\SupplementRef{sm:#1}}
\endgroup
\end{document}